\documentclass[aps,prc,superscriptaddress,twoside,twocolumn,nofootinbib,showpacs]{revtex4}

\usepackage[pdftex]{graphicx,color}
\usepackage{natbib,subfigure}
\usepackage{amsmath,amssymb,latexsym}
\usepackage{changes}

\usepackage{pgf}

\usepackage{bm}

\usepackage{subfigure}

\usepackage{tabularx}
\usepackage{url}

\usepackage[colorlinks=true,linkcolor=blue,citecolor=blue]
{hyperref}

\allowdisplaybreaks

\usepackage{footnote}

\usepackage{hyperref}

\usepackage{marvosym}
\usepackage{bbold}
\usepackage{textgreek}

\begin{document}
\title{Universal correlations in the nuclear symmetry energy, slope parameter, and curvature}

\author{Jeremy~W. Holt}
 \email{holt@physics.tamu.edu}
\affiliation{Cyclotron Institute, Texas A\&M University, College Station, TX 77843, USA}
\affiliation{Department of Physics and Astronomy, Texas A\&M University, College Station, TX 77843, USA }

\author{Yeunhwan Lim}
 \email{ylim@tamu.edu}
\affiliation{Cyclotron Institute, Texas A\&M University, College Station, TX 77843, USA}

\date{\today}

\begin{abstract}
From general Fermi liquid theory arguments, we derive correlations among the symmetry energy ($J$), 
its slope parameter ($L$), and curvature ($K_{\rm sym}$) at nuclear matter saturation density. 
We argue that certain properties of these correlations do not depend on details of the
nuclear forces used in the calculation. We derive as well a global parametrization 
of the density dependence of the symmetry energy that we show is more reliable,
especially at low densities, than the usual Taylor series expansion around saturation
density. 
We then benchmark these predictions against explicit results from chiral effective field theory.
\end{abstract}

\maketitle









The nuclear isospin-asymmetry energy, which characterizes the energy cost of converting protons
into neutrons in an interacting many-body system, is an 
important organizing concept linking the properties of atomic nuclei to the structure and dynamics
of neutron stars.
In particular the isospin-asymmetry energy 
governs the proton fraction of dense matter in beta equilibrium, the thickness of neutron star crusts, 
and the typical radii of neutron stars 
\cite{lattimer00,lattimer01,steiner05,lattimer12,steiner12,gandolfi12}.
For these reasons the nuclear isospin-asymmetry energy is a primary focus of experimental investigations
at current and next-generation rare-isotope facilities such as the 
Radioactive Isotope Beam Factory (RIBF), the Facility for Antiproton and Ion Research (FAIR), 
and the Facility for Rare Isotope Beams (FRIB).

In recent years, theoretical \cite{xu10,kortelainen10,hebeler10,gandolfi12,holt17prc} and
experimental \cite{shetty07,tsang09,centelles09,chen10,carbone10} studies have reduced the
uncertainties on the isospin-asymmetry energy at and below the density scales of normal nuclei, but more 
challenging is to derive constraints at the higher densities reached in the cores of neutron stars. 
Given the experimental difficulties of creating and studying high-density, low-temperature matter 
in the lab, an alternative strategy has been to extract the coefficients in the Taylor series expansion of the 
isospin-asymmetry energy about nuclear matter saturation density. 
For instance, the slope parameter has been shown to correlate strongly with neutron skin thicknesses 
in nuclei \cite{centelles09,Roca2011,vina2014}, nuclear electric dipole 
polarizabilities~\cite{reinhard2010,tamii2011,piekarewicz2012,Roca2015a,hashi2015,tonchev2017}, 
and the difference in charge radii of mirror nuclei~\cite{Alex2017,Yang2018}. 
Determining the isospin-asymmetry energy 
curvature is more challenging~\cite{vidana2009,ducoin2011,santos2014}
with larger associated uncertainties.

A feature observed in many experimental and theoretical investigations
is a nearly linear correlation between the value of the isospin-asymmetry energy at nuclear
matter saturation density, its slope, and curvature (for a recent comprehensive analysis,
see Ref.\ \cite{tews17}). 
In Ref.\ \cite{holt17prc} it was shown that even chiral nuclear potentials at next-to-leading 
order (NLO), which are rather simplistic and contain no three-body forces, exhibit a
correlation slope consistent with previous microscopic calculations at N2LO and N3LO in the chiral 
expansion. This suggests that certain aspects of the correlation are ultimately associated with
low-energy physics well described even at next-to-leading order in the chiral expansion. 
In the present work we will demonstrate that this is indeed the case and that the slope of the
correlation can be derived without referring to detailed properties of the nucleon-nucleon potential.
The overall scale is then set by a few constants that can in principle be extracted from the 
properties of low-density homogeneous matter.
We also show that the same arguments can be used to derive the slope of the correlation 
between the symmetry energy and its curvature at nuclear matter saturation density.

We take as a starting point the nuclear symmetry energy $S(\rho)$, which is defined 
as the difference in the energy 
per nucleon between neutron matter and symmetric nuclear matter at a given density:
\begin{equation}\label{eq:symi}
S(\rho) = \frac{E}{N}(\rho,\delta_{np}=1) -  \frac{E}{N}(\rho,\delta_{np}=0),
\end{equation}
where $\rho$ is the total baryon number density, $N$ is the total baryon number,
and $\delta_{np}=(\rho_n-\rho_p)/(\rho_n+\rho_p)$ is the isospin asymmetry parameter.
A Maclaurin expansion of the nuclear equation of state around symmetric nuclear matter
\begin{equation}
\frac{E}{N}(\rho,\delta_{np}) = \sum_{n=0}^\infty A_{2n}(\rho) \delta_{np}^{2n}
\end{equation}
is in general \cite{kaiser15,wellenhofer16} noncovergent due to the appearance of 
nonanalytic logarithm terms that appear beyond a mean field description 
of the nuclear equation of state:
\begin{eqnarray}
&&\hspace{-.4in}\frac{E}{N}(\rho,\delta_{np}) = A_0(\rho) + S_2(\rho) \delta_{np}^2 \nonumber \\
&&+ \sum_{n=2}^\infty (S_{2n}+L_{2n}\ln | \delta_{np} |)\, \delta_{np}^{2n}.
\end{eqnarray}
However, all contributions beyond the quadratic term $S_2$, which we call the isospin-asymmetry
energy, have been shown to be small at low temperatures when realistic nuclear forces are 
employed \cite{bombaci91,lee98,zuo99,zuo02,frick05,PhysRevC.89.025806,wellenhofer16}. 
It is therefore a good approximation to identify the nuclear symmetry
energy $S(\rho)$ with the isospin-asymmetry energy $S_2(\rho)$. 
In practice it is the latter quantity that can be inferred from laboratory measurements of 
finite nuclei. 

\begin{figure}[t]
\centering
\includegraphics[scale=0.34]{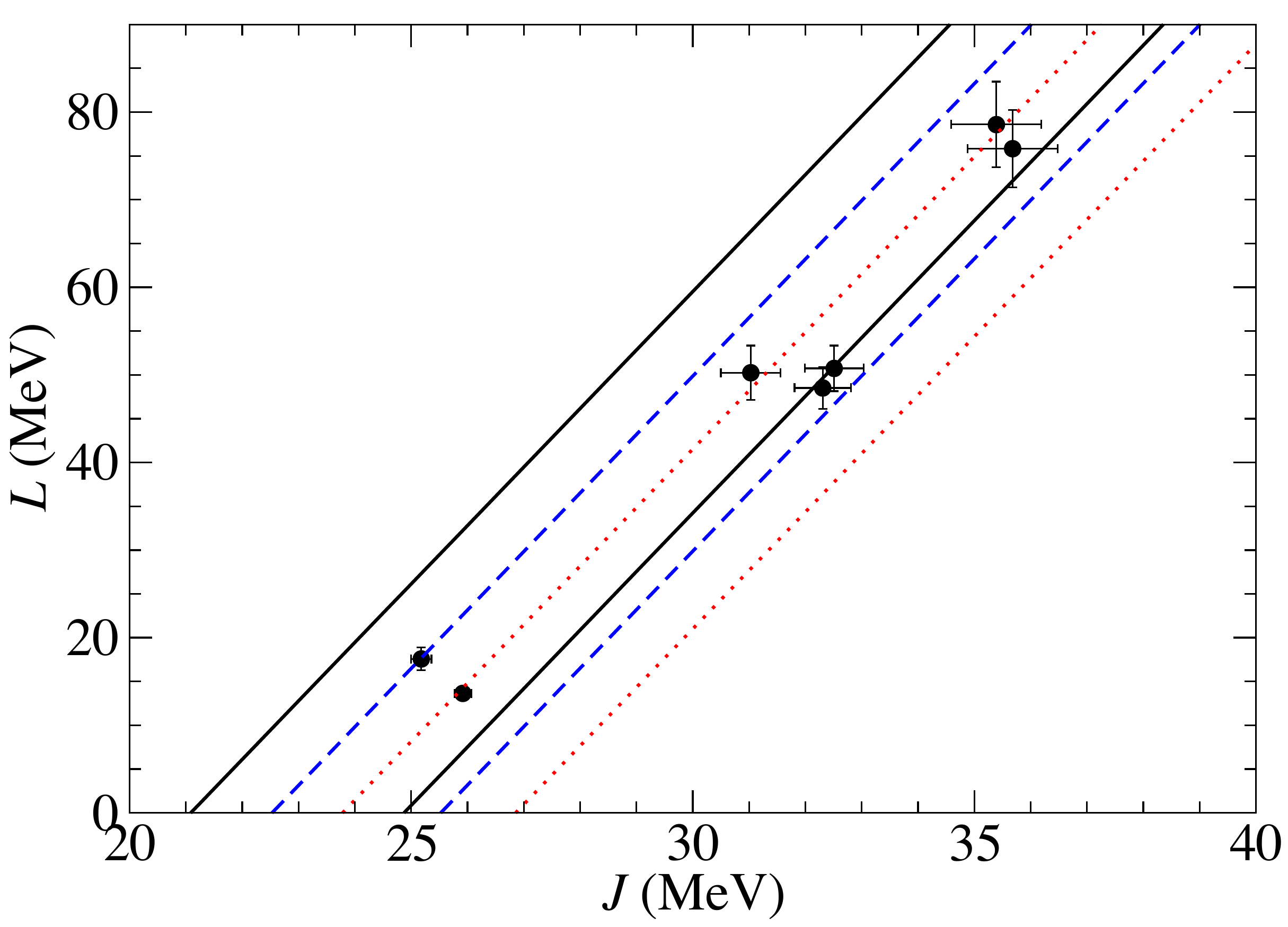}
\caption{(color online) Symmetry energy $J$ vs.\ its slope parameter $L$ from the seven
chiral nuclear potentials considered in this work. The bands are obtained by keeping all terms up
to $a_0$ (red dotted), $a_1$ (blue dashed), and $a_2$ (black solid) in Eq.\ (\ref{newLc}).}
\label{svsl}
\end{figure}

It is common to expand the density dependence of the isospin-asymmetry energy 
in a Taylor series around nuclear saturation:
\begin{equation}
S_2(\rho) = J + L \left(\frac{\rho-\rho_0}{3\rho_0} \right)
+ \frac{1}{2}K_{\rm sym}\left(\frac{\rho-\rho_0}{3\rho_0} \right)^2 + \cdots\,,
\label{eq:symdef}
\end{equation}
where $J \equiv S_2(\rho_0)$.
The parameters $L$ and $K_{\rm sym}$ can then be extracted from
properties of finite nuclei \cite{PhysRevC.85.035201}, but their uncertainties are much larger than 
that associated with the isospin-asymmetry energy. For this reason correlations between the nuclear 
symmetry energy, the slope parameter, and curvature are routinely investigated within
a range of theoretical models \cite{PhysRevC.80.014322,DANIELEWICZ200936,vidana2009,xu10,ducoin2011,santos2014,tews17}.

In Fermi liquid theory the nuclear isospin-asymmetry energy is rigorously defined at
all densities in terms of the isotropic component $f_0^\prime$ of the isovector-scalar quasiparticle 
interaction according to
\begin{equation}
S_2(\rho) = \frac{k_F^2}{6m^*}\left (1+\frac{2m^*k_F}{\pi^2}f_0^\prime \right),
\label{flps}
\end{equation}
where $k_F$ is the Fermi momentum and the nucleon effective mass $m^*$ is
related to the Fermi liquid parameter $f_1$ through
\begin{equation}
\frac{1}{m^*} = \frac{1}{m}-\frac{2k_F}{3\pi^2}f_1.
\end{equation}
These Fermi liquid parameters are obtained by performing a Legendre polynomial 
decomposition of the central quasiparticle interaction
\begin{eqnarray}
&&\hspace{-.5in} {\cal F}(\vec p_1, \vec p_2) 
= \sum_l \left ( f_l + g_l \vec \sigma_1 \cdot \vec \sigma_2 + f_l^\prime \vec \tau_1 \cdot \vec \tau_2 
\right . \nonumber \\
&& \hspace{.3in}\left . + g_l^\prime \vec \sigma_1 \cdot \vec \sigma_2 
\vec \tau_1 \cdot \vec \tau_2 \right ) P_l (\cos \theta),
\end{eqnarray}
in the relative angle $\cos \theta = \hat p_1 \cdot \hat p_2$.
Expanding Eq.\ (\ref{flps}) we obtain
\begin{equation}
S_2(\rho) = \frac{k_F^2}{6m} + \frac{k_F^3}{9\pi^2} \left [ 3f_0^\prime (\rho) - f_1 (\rho) \right ].
\label{sflt}
\end{equation}

The relationship in Eq.\ (\ref{sflt}) can be used to derive a correlation between the symmetry
energy slope parameter $L$ and the symmetry energy $J$ at nuclear matter saturation density:
\begin{eqnarray}
&& \hspace{-.3in}L = 3 \rho_0 \left . \frac{dS_2}{d\rho} \right |_{\rho_0} = 3 J - S_0
\nonumber \\
&&\hspace{.6in}+ \frac{\rho_0}{6} \left . \left (
3k_F\frac{df_0^\prime}{dk_F}-k_F\frac{df_1}{dk_F} \right ) \right |_{k_F^0},
\label{expL}
\end{eqnarray}
where $\rho_0 = 0.16$\,fm$^{-3}$ is nuclear matter saturation density, $k_F^0$ is the 
Fermi momentum at saturation density, and
$S_0 \equiv (k_F^0)^2/(6m)$ is the noninteracting part of the isospin-asymmetry energy
at saturation density.
To study the additional correlation between $L$ and $J$ associated with the density
derivative terms on the right-hand side of Eq.\ (\ref{expL}), 
we perform a Taylor series expansion of the quantity
$3 f_0^\prime - f_1$ around a small reference density set by $k_F = k_r$:
\begin{equation}
3f_0^\prime(k_F) - f_1(k_F) = a_0 + a_1\, \beta + \frac{1}{2} a_2\, \beta^2 + \cdots ,
\label{tayexp}
\end{equation}
where $\beta = (k_F-k_r)/k_r$.

In principle the choice of reference Fermi momentum $k_r$ is arbitrary, but we note 
several constraints. First, logarithmic contributions to the isospin-asymmetry 
energy of the form $\log(1+4k^2_F/m_\pi^2)$ arise from the one-pion-exchange 
Fock diagram \cite{kaiser02} and formally require that $k_r \gtrsim 0.9$\,fm$^{-1}$ 
in order for the Taylor series to be convergent at nuclear matter saturation density. 
Second, $k_r$ should be 
small enough that a reliable calculation of the 
Fermi liquid parameters $f_0^\prime(k_r)$ and $f_1(k_r)$ may be achieved within chiral 
effective field theory whose formal expansion parameter at the reference density would
be $k_r / \Lambda_\chi$, where $\Lambda_\chi \sim 500$\,MeV is a typical 
momentum-space cutoff scale in realistic chiral nucleon-nucleon potentials.
The effect of the less certain three-body forces, which start contributing to the homogeneous
matter equation of state at a density $\rho \simeq 0.03$\,fm$^{-3}$ \cite{PhysRevC.82.014314,wlazlowski14,tews16},
should also be minimized. From these considerations we choose the reference Fermi momentum 
$k_r = 0.9$\,fm$^{-1} = 175$\,MeV, which satisfies $( k_F^0 - k_r ) / k_r \simeq 0.5$, where $k_F^0$ 
is the Fermi momentum of nuclear matter
at saturation density. We will show later that this value, which corresponds to the density 
$\rho \simeq 0.05$\,fm$^{-3}$, is large enough that keeping the first three terms in the 
Taylor series expansion in Eq.\ (\ref{tayexp}) 
provides a good description of the isospin-asymmetry energy somewhat above nuclear saturation density,
even though formally the series fails to converge in that regime.

Only the terms in Eq.\ (\ref{tayexp}) proportional to $a_1$ and $a_2$ result in additional 
correlations between $L$ and $J$. Our strategy is therefore to absorb the dependence of the
$L$ vs.\ $J$ correlation on the choice of nuclear interaction into the coefficient $a_0$, 
which we expect to be very similar for different nuclear force models.
Combining Eqs.\ (\ref{tayexp}) and (\ref{expL}) and
defining $\gamma \equiv 2k_F^2/(k_F^2-k_r^2)$, we find the unique solution
\begin{eqnarray}
&& \hspace{-.4in} L = (3 +\gamma)J - (1+\gamma)S_0 \nonumber \\ 
&& - \gamma \frac{\rho_0}{6} \left ( a_0 - \eta_1 a_1 + \eta_1 a_2 \right ),
\label{newLc}
\end{eqnarray}
where $\eta_1 = (\beta + 1)\gamma^{-1}-\beta$. This rearrangement
simultaneously minimizes the residual importance of $a_1$ and
$a_2$ after we have included their effect on the $L$ vs.\ $J$ correlation
through the term $(3+\gamma)J$. For instance, if $k_r = 0.9$\,fm$^{-1}$, then 
$\gamma \simeq 3.7$ and $\eta_1 \simeq -0.08$. We therefore expect $a_0$ to give the dominant 
contribution to the last term on the right-hand side of Eq.\ (\ref{newLc}). 

\begin{figure}[t]
\centering
\includegraphics[scale=0.33]{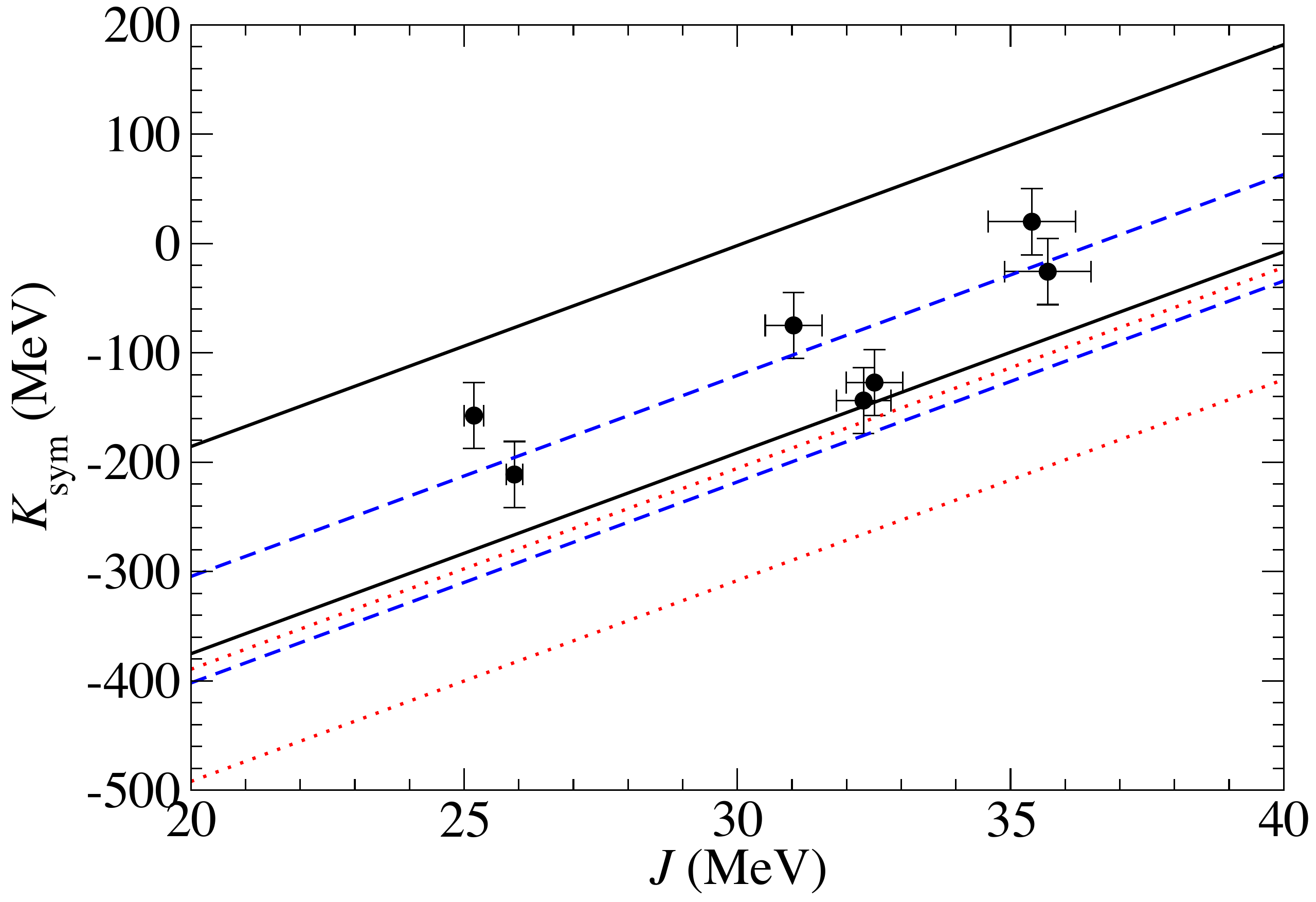}
\caption{(Color online) Symmetry energy $J$ vs.\ its curvature parameter $K_{\rm sym}$ from the seven
chiral nuclear potentials considered in this work. The bands are obtained by keeping all terms up
to $a_0$ (red dotted), $a_1$ (blue dashed), and $a_2$ (black solid) in Eq.\ (\ref{newKc}).}
\label{svsk}
\end{figure} 

In Fig.\ \ref{svsl} we show the values of the symmetry energy and its slope
parameter at nuclear matter saturation density for a set of seven realistic nuclear potentials
\cite{entem03,coraggio13,coraggio14,sammarruca15}
at different orders in the chiral expansion and different values of the momentum-space
cutoff $\Lambda$:
\{N1LO\_450, N1LO\_500, N2LO\_450, N2LO\_500, N3LO\_414, N3LO\_450, N3LO\_500\}.
Since the equation of state of pure neutron matter is well converged in many-body 
perturbation theory up to nuclear matter saturation density using modern chiral effective 
field theory forces with $\Lambda \lesssim 500$\,MeV \cite{holt17prc}, we compute in Fig.\ \ref{svsl} the symmetry energy
assuming a symmetric nuclear matter saturation density $\rho=0.155-0.165$\,fm$^{-3}$
and a saturation binding energy of $E/N = 16$\,MeV. The error bars on the data points
arise from varying the saturation density within the above range. We can also extract the 
isospin-asymmetry energy directly from a calculation of the quasiparticle interaction 
in Fermi liquid theory starting from chiral two- and three-body forces as described in
Refs.\ \cite{holt11npa,holt12npa,PhysRevC.87.014338,holt18prc}.
In Fig.\ \ref{svsl} we show as well the correlation bands obtained by keeping in Eq.\ (\ref{newLc})
only the $a_0$ term (red-dotted lines), the $a_0$ and $a_1$ terms (blue-dashed lines),
and all three terms $a_0$, $a_1$, and $a_2$ (black solid lines) by fitting the density-dependent
isospin-asymmetry energy from Eqs.\ (\ref{sflt}) and (\ref{tayexp})
over the range $0.85$\,fm$^{-1} < k_F < 1.8$\,fm$^{-1}$
for each of the seven chiral nuclear forces considered in this work. We see that indeed the
model-independent prediction for the slope of the $J$ vs.\ $L$ correlation, $m_L \simeq 6.7$, 
is well satisfied by all N1LO, N2LO, and N3LO chiral nuclear forces. We also observe that the size of the 
uncertainty band is set already from the uncertainties in $a_0$ and that the inclusion of $a_1$ and
$a_2$ leads primarily to a shift of the band.

\begin{figure}[t]
\centering
\includegraphics[scale=0.34]{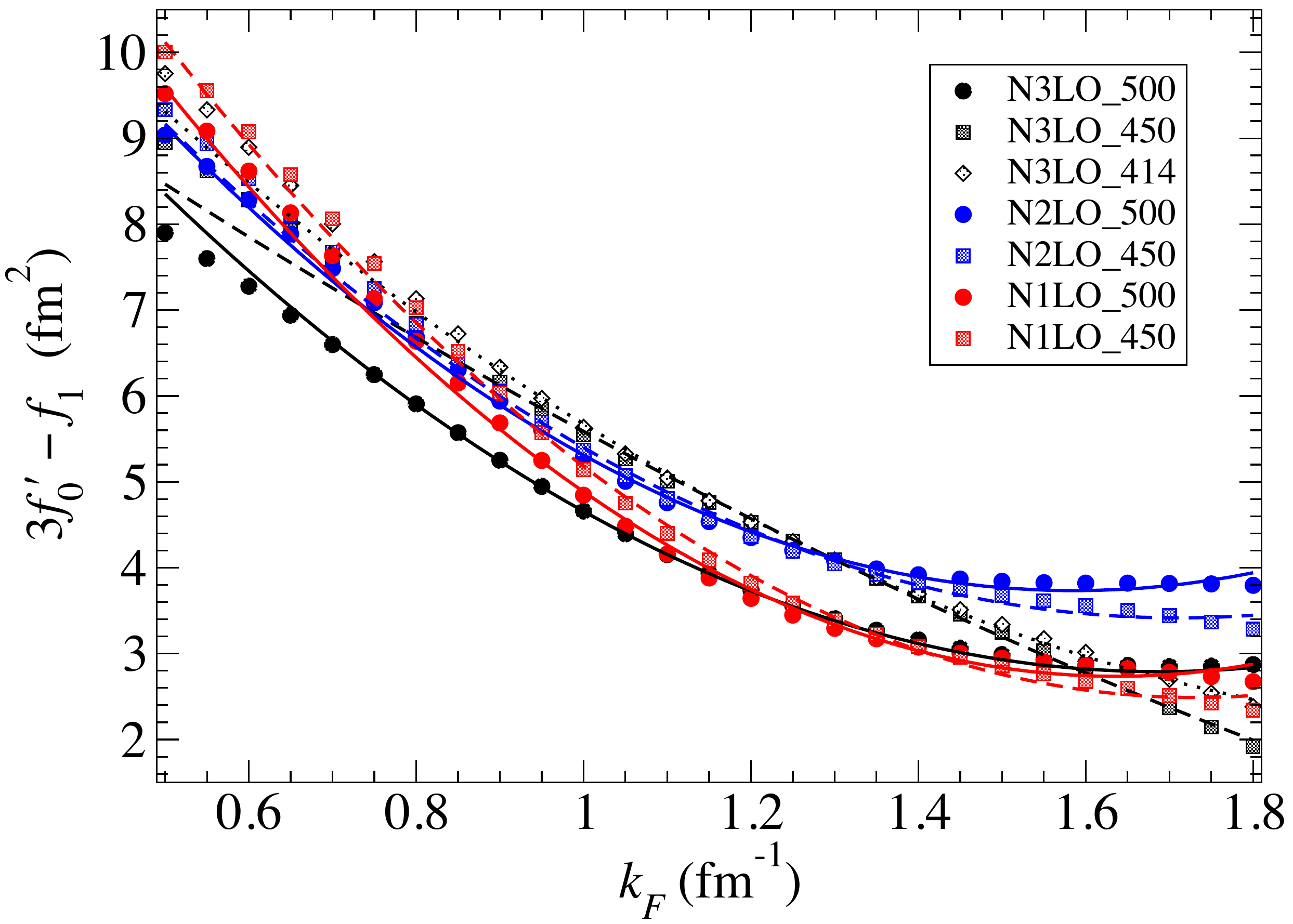}
\caption{(Color online) Combination of the Fermi liquid parameters associated with the isospin-asymmetry
energy as a function of Fermi momentum. The solid lines are fit functions of the form given in 
Eq.\ (\ref{tayexp}) over the range $0.85$\,fm$^{-1} < k_F < 1.8$\,fm$^{-1}$.}
\label{fitall}
\end{figure}

\begin{figure*}[t]
  \centering
  \subfigure{\includegraphics[scale=0.45]{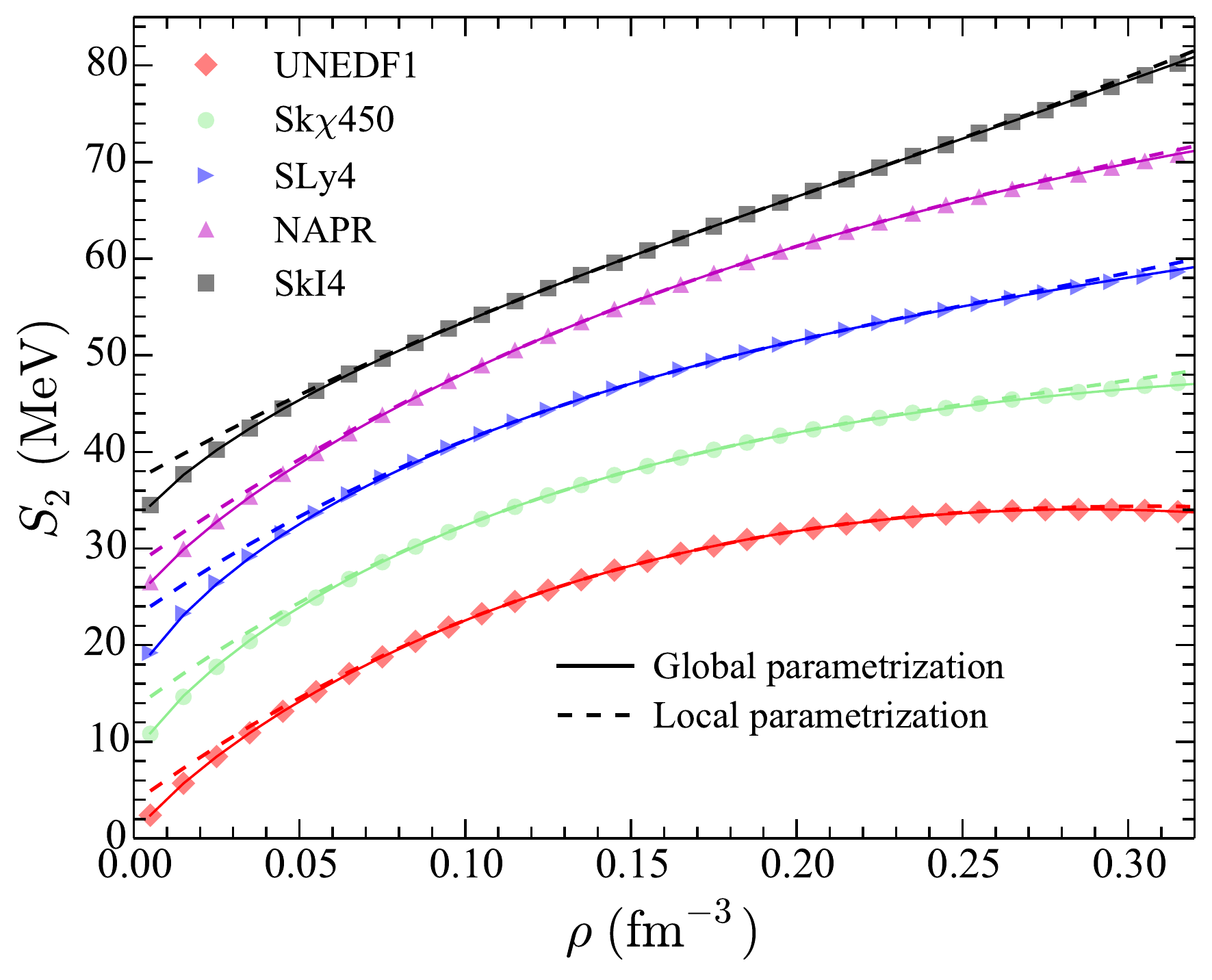}}\quad
  \subfigure{\includegraphics[scale=0.363]{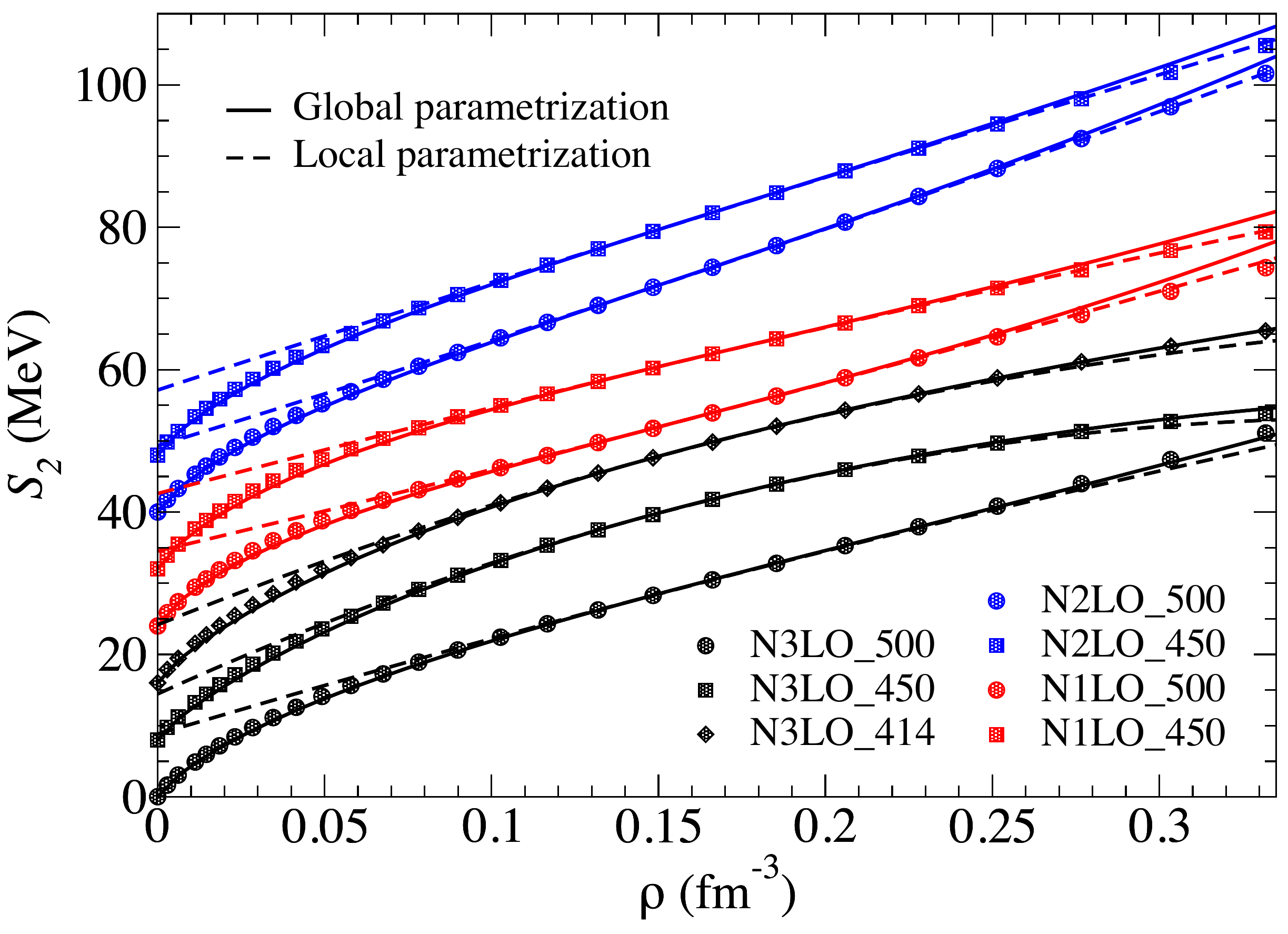}}
\caption{(Color online) Left: nuclear isospin-asymmetry energy as a function of density from 
various Skyrme force models. Right: nuclear isospin-asymmetry energy as a function of density 
from various chiral nuclear forces. Note that in both figures we have introduced a vertical offset to 
better distinguish the different curves.}
\label{fig:symfor}
\end{figure*} 

We can extend this analysis to derive an additional correlation between the symmetry energy
incompressibility $K_{\rm sym}$ and $J$:
\begin{eqnarray}
&&\hspace{-.2in} K_{\rm sym} = 9 \rho_0^2 \left . \frac{d^2S_2}{d\rho^2} \right |_{\rho_0} 
= 4 L - 12J + 2S_0 \nonumber \\ 
&& \hspace{.7in}+ \frac{\rho_0}{6} \left . 
\left ( 3k_F^2\frac{d^2f_0^\prime}{dk_F^2}-k_F^2\frac{d^2f_1}{dk_F^2} \right ) \right|_{k_F^0}.
\label{expL}
\end{eqnarray}
By again minimizing the explicit dependence on $a_1$ and $a_2$ we obtain the unique solution
\begin{eqnarray}
&& \hspace{-.4in} K_{\rm sym} = 5 \gamma J - (5\gamma+2)S_0 \nonumber \\ 
&& \hspace{.2in} - 5\gamma \frac{\rho_0}{6} \left ( a_0 - \eta_2 a_1 + \eta_2 a_2 \right ),
\label{newKc}
\end{eqnarray}
where $\eta_2 = 4(\beta + 1)\gamma^{-1}-5\beta$. Now for 
$k_r=0.9$\,fm$^{-1}$, we obtain $\eta_2 \simeq -0.16$ and therefore we expect the residual 
importance of $a_1$ and $a_2$ to be greater than in the $L$ vs.\ $J$ correlation.
Nevertheless, we have derived a second universal slope parameter
$m_K = 5 \gamma \simeq 18.4$ for the $J$ vs.\ $K_{\rm sym}$ correlation that is
independent of details of the nuclear force employed.

In Fig.\ \ref{svsk} we show the values of the symmetry energy $J$ and curvature $K_{\rm sym}$
at nuclear saturation density using the same set of chiral potentials. The largest source of uncertainty
(represented by the error bars on the correlation points)
is the assumed value the symmetric nuclear matter incompressibility $K_0 = 220-260$\,MeV
\cite{youngblood99,shlomo06}. We show as well the correlation bands obtained by keeping in 
Eq.\ (\ref{newKc}) only the $a_0$ term (red-dotted lines), the $a_0$ and $a_1$ terms (blue-dashed lines),
and all three terms $a_0$, $a_1$, and $a_2$ (black solid lines). 
Again we see that the slope of the correlation is well determined
from our model-independent analysis and that the overall spread in the uncertainty band is set
with the inclusion of only the $a_0$ term in Eq.\ (\ref{newKc}).

In Fig.\ \ref{fitall} we show the Fermi momentum dependence of the quantity 
$3f_0^\prime(k_F) - f_1(k_F)$ for the seven chiral potentials considered in this work. 
We note that the N3LO\_500 chiral
potential exhibits the poorest convergence in many-body perturbation theory, which may
account for its different low-density behavior. When
fitting the theoretical results with the functional form given in Eq.\ (\ref{tayexp}), we include
only the points for which $k_F > 0.85$\,fm$^{-1}$. For smaller values of the 
Fermi momentum, we found that third-order perturbative contributions (not included explicitly 
in this work) become important. We see that indeed the functional form in Eq.\ (\ref{tayexp})
is able to well reproduce the Fermi momentum dependence of the Fermi liquid parameters
up to $k_F \simeq 1.8$\,fm$^{-1}$. We find for the values of the expansion coefficients: 
$a_0 = 5.88 \pm 0.35$\,fm$^{2}$, $a_1 = -6.04 \pm 0.91$\,fm$^{2}$, and
$a_2 = 6.08 \pm 2.48$\,fm$^{2}$ averaged over the seven chiral potentials. Combined with 
the expansion parameter $\beta = 0.5$, we see that the explicit calculations suggest a 
convergent series at nuclear matter saturation density. 

Finally we observe that the ansatz for the density dependence 
of the isospin-asymmetry energy in Eqs.\ (\ref{sflt}) and (\ref{tayexp})
produces the first four terms in the general expansion
\begin{equation}
\label{eq:symant}
S_2(\rho) = \sum_{i=0}^N b_i \left ( \frac{\rho}{\rho_0} \right )^{(i+2)/3}.
\end{equation}
Combining Eq.~\eqref{eq:symant} with the definition of the symmetry energy parameters in 
Eq.~\eqref{eq:symdef},
we obtain
\begin{equation}
\begin{aligned}
\label{eq:symcoeff}
b_0 & = S_0,  \\
b_1 & = \frac{1}{2}K_{\rm sym} - 3L + 10J - 3S_0,  \\
b_2 & =  - K_{\rm sym} +  5L - 15J + 3S_0, \\
b_3 & = \frac{1}{2}K_{\rm sym} -2L + 6J -S_0\,.
\end{aligned}
\end{equation}
These expressions are of course independent of the choice of reference Fermi momentum $k_r$.
In Fig.\ \ref{fig:symfor} we show the accuracy of this global parametrization (solid lines) compared 
to the normal Taylor series expansion about nuclear saturation density (dashed lines) in
Eq.\ (\ref{eq:symdef}). We 
show results for both a representative set of Skyrme mean field models (on the left) as well as 
the seven chiral interactions (on the right). We note that a vertical offset is used in order to 
separate the curves. In general the Taylor expansion in Eq.\ (\ref{eq:symdef}) does not 
reproduce well the low-density behavior of the isospin-asymmetry energy, especially in the 
case of chiral nuclear interactions.
Each Skyrme force model contains a density-dependent interaction with
corresponding contribution to the symmetry energy
\begin{equation}
S_{2d}(\rho) = 
-\frac{t_3}{24}\left(\frac{1}{2} + x_3 \right)\rho^{1+\epsilon_1}
-\frac{t_4}{24}\left(\frac{1}{2} + x_4 \right)\rho^{1+\epsilon_2},
\end{equation}
where $t_i$ and $x_i$ are independent of density.
For example, UNEDF1~\cite{unedf1}, 
SLy4~\cite{CHABANAT1998231}, 
NAPR~\cite{steiner05}, and
SkI4~\cite{ski4}
have the values $\epsilon_1= \{0.27, 1/6, 0.1441, 1/4\}$,
while the Sk$\chi$450~\cite{lim2017a} interaction has the
two parameters $\epsilon_1=1/3, \epsilon_2=1$. 
Thus, the fitting function in Eqs.~\eqref{eq:symant} and \eqref{eq:symcoeff} 
is quite flexible, and we suggest that it may provide a more useful global parametrization of the 
nuclear isospin-asymmetry energy.

In summary we have derived correlations between the nuclear isospin-asymmetry energy, 
its slope parameter, and curvature within a Fermi liquid theory description of nuclear matter.
We derived universal slope parameters for the $J$ vs.\ $L$ and $J$ vs.\
$K_{\rm sym}$ correlations that are nearly independent of details of the nuclear interaction. We have assumed
only that the quasiparticle interaction in nuclear matter at the low density scale set by 
$k_F = k_r =0.9$\,fm$^{-1}$ should be well described by any realistic nucleon-nucleon potential
fitted to scattering phase shifts. Future efforts to reduce the theoretical uncertainties on 
the Fermi liquid parameters at this low-density scale may then give more stringent constraints
on the symmetry energy correlation curves.

\vspace{.1in}
\noindent {\bf Acknowledgments}

Work supported by the National Science Foundation under grant 
No.\ PHY1652199.




\bibliographystyle{apsrev4-1}

%

\end{document}